\documentclass[preprint,aps,amsmath,amssymb,12pt]{revtex4}

\usepackage{epsfig}
\usepackage{slashed}
\usepackage{graphicx}
\usepackage{multirow,color}
\usepackage{amsmath}
\usepackage{float}
\usepackage{diagbox}
\usepackage{CJK}
\usepackage{color}
\usepackage{xcolor}
\usepackage{times}
\usepackage{subfigure}
\usepackage{bm}
\usepackage{braket}
\usepackage{booktabs}
\usepackage{array}
\usepackage[mathscr]{euscript}
\usepackage{caption}
\usepackage[justification=centering]{caption}
\usepackage{makecell}

\makeatletter

\newcommand{\Rmnum}[1]{\expandafter\@slowromancap\romannumeral #1@}
\makeatother
\graphicspath{{fig/}}
\begin{document}
\title{Searching for singlet vector-like leptons via pair production at ILC}
\author{Chong-Xing Yue$^{1,2}$}
\thanks{cxyue@lnnu.edu.cn}
\author{Yue-Qi Wang$^{1,2}$}
\thanks{wyq13889702166@163.com}
\author{Han Wang$^{1,2}$}
\thanks{wangwanghan1106@163.com}
\author{Yi-Hang Wang$^{1,2}$}
\thanks{wyh15841047016@163.com}
\author{Si Li$^{1,2}$}
\thanks{lisi021777@163.com}

\affiliation{
$^1$Department of Physics, Liaoning Normal University, Dalian 116029, China\\
$^2$Center for Theoretical and Experimental High Energy Physics, Liaoning Normal University, China
}

\begin{abstract}
Vector-like leptons (VLLs) as one kind of the most intriguing particles, have been widely concerned in several extensions of the Standard Model (SM). In this work, we explore the discovery potential of VLLs via  pair production in the context of models that satisfy asymptotic safety at the International Linear Collider (ILC). The expected sensitivities of the ILC with the center of mass energy $\sqrt{s} =$ 1 TeV and the integrated luminosity $\mathcal{L}$ = 1 ab$^{-1}$ to the parameter space of this kind of VLL models are derived. The results we obtained show that, for the VLL with mass $M_F$ in the region of $M_{F}\in$ [101GeV, 463GeV], the Yukawa coupling $\kappa $ can be  as low as $\kappa \in$ [0.032, 0.098].
\end{abstract}


\maketitle
\section{Introduction}
Although the discovery of the Higgs boson~\cite{ATLAS:2012yve,CMS:2012qbp} indicates the completion of the Standard Model (SM) of elementary particles, there still remain some unresolved issues that require to make clear. For example, it can not deal with the existence of gauge hierarchy~\cite{Feng:2013pwa} and neutrino masses~\cite{Gonzalez-Garcia:2007dlo}.  Moreover, mass and mixing patterns of the SM leptons~\cite{ParticleDataGroup:2022pth} cannot be stated clearly. So, large number of new physics models beyond the SM have been proposed and  these models predict  existence of new particles. The vector-like fermions including vector-like quarks (VLQs) and vector-like leptons (VLLs) as one kind of the most intriguing particles have recently attracted widespread attention both in theory and experiment. It is well known that vector-like quarks have been studied extensively in literatures, for example see the classical papers~\cite{He:2014ora,Wang:2013jwa,He:2001fz,He:1999vp} and recent review article~\cite{Alves:2023ufm}. However, it is important to note that, from the point of view of particle phenomenology, vector-like leptons have the same status as vector-like quarks.

So far, there are many extensions of the SM that predict the existence of VLLs, such as composite models~\cite{Panico:2015jxa,Cacciapaglia:2020kgq}, left-right symmetric models~\cite{Pati:1974yy,Mohapatra:1974gc,Senjanovic:1975rk,Mohapatra:1977mj}, supersymmetric models~\cite{Martin:2009bg,Endo:2011mc,Araz:2018uyi}, and grand unified theories~\cite{Morais:2020odg,Morais:2020ypd}. The extensively searches of the VLLs with masses from a few GeV to TeV consist of an important component of the experimental and theoretical expectations. The parameter space and the variety of mass ranges of VLLs have been limited by different high energy collider experiments. An early study for VLLs at the Large Electron-positron (LEP) excluded the heavy leptons with mass up to 101.2 GeV~\cite{L3:2001xsz}. At the Large Hadron Collider (LHC), CMS excluded VLLs transforming as singlets under $SU(2)_L$ in the range from 125 to 150 GeV at 95\% CL~\cite{CMS:2022nty}. For doublet vector-like $\tau$ lepton extension of the SM, a recent ATLAS study based on 139 fb$^{-1}$ at the 13 TeV LHC excluded VLLs in the mass range 130-900 GeV at 95\% CL~\cite{ATLAS:2023sbu}. Using the CMS search based on 77.4 fb$^{-1}$ at the 13 TeV LHC, the authors of Ref.~\cite{Bissmann:2020lge} have found that flavorful $SU(2)_L$ singlet VLLs are excluded for mass below around 300 GeV and the doublet VLLs are excluded below around 800 GeV. Additionally, other noteworthy discussions regarding VLLs in the context of different new physics scenarios have been made in Refs.~\cite{Halverson:2014nwa,Schwaller:2013hqa,Ishiwata:2013gma,Dermisek:2014cia,Dermisek:2015oja,Chen:2016lsr,
Xu:2018pnq,Zheng:2019kqu,Freitas:2020ttd,Dermisek:2021ajd,Dermisek:2020cod,Kawamura:2021ygg,Raju:2022zlv,Kawamura:2023zuo,Cao:2023smj,Liu:2010ze}.

At present, most of the research is based on different VLL models or in model-independent way. In this work, we explore the discovery potential of VLLs at the International Linear Collider (ILC) in the context of the models that satisfy asymptotic safety~\cite{Litim:2014uca,Bond:2018oco,Bond:2019npq,Bond:2017lnq,Bond:2017sem,Bond:2016dvk,Bednyakov:2023fmc,Litim:2023tym} and introduce new matter fields and Yukawa couplings~\cite{Bissmann:2020lge,Hiller:2020fbu,Hiller:2019tvg,Hiller:2019mou}. The new models called VLS models in this paper, contain VLLs and new scalars, which tame the UV behavior~\cite{Gross:1973id} of the Standard Model towards the Planck scale or beyond. After spontaneous symmetry breaking, the Z, W and Higgs boson couplings will be modified due to the mixing of VLLs, new scalars with the SM leptons, Higgs boson, respectively. In this paper, we focus on the couplings of VLLs to neutrino and W boson and investigate the possibility of searching for the VLLs at the 1 TeV ILC for an integrated luminosity of $\mathcal{L}=1$ ab$^{-1}$~\cite{ILC:2019gyn,Bambade:2019fyw}.

The paper is structured as follows. The interaction lagrangian of VLLs with the SM particles is introduced in section II. Detailed analysis for the possibility of probing VLLs via pair production at the ILC is provided in section III. In section IV, we summarize the main results about the projected sensitivity of the ILC to the couplings of VLLs with W bosons for different values of the VLL mass.

\section{Effective interactions of VLLs}
The VLS models~\cite{Hiller:2020fbu,Hiller:2019tvg,Hiller:2019mou} as one kind of the new VLL models predict the existence of the three generations of VLLs $\psi_{L,R}$, which may be $SU(2)_L$ singlets with hypercharge $Y = -1$ and $SU(2)_L$ doublets with $Y = -\frac{1}{2}$ corresponding the singlet and doublet VLS models, respectively. Two models also involve complex scalars $S_{ij}$ as singlets under the SM gauge interactions, where $i,j = 1, 2, 3$ are two flavor indices. The new Yukawa lagrangians of the VLS models read
\begin{eqnarray}
\begin{split}
\mathcal{L}_{\text{Y}}^{\text{singlet}}
	=
    &-\kappa\bar{L}H\psi_{R} -\kappa'\bar{E}S^{\dagger }\psi_{L}-y\bar{\psi}_{L}S\psi_{R}+h.c.,
\end{split}
\end{eqnarray}
\begin{eqnarray}
\begin{split}
\mathcal{L}_{\text{Y}}^{\text{doublet}}
	=
    &-\kappa\bar{E}H^{\dagger }\psi_{L} -\kappa'\bar{L}S\psi_{R}-y\bar{\psi}_{L}S\psi_{R}+h.c..
\end{split}
\end{eqnarray}
Where E, L and H denote the SM  singlet, doublet leptons and Higgs boson, respectively. On account of $SU(3)$-flavor symmetries of VLS models, each lepton fulfills flavor-conservation and the new Yukawa couplings $y, \kappa, \kappa'$ become single couplings, instead of being tensors~\cite{Hiller:2020fbu}.

After spontaneous symmetry breaking, the effective interaction among VLLs with the new scalar particles and SM particles in the singlet VLS model is given by Ref.~\cite{Bissmann:2020lge}.
\begin{eqnarray}
\begin{split}
\mathcal{L}_{\text{int}}^{\text{singlet}}
	=
    &-e\bar{\psi}\gamma^{\mu}\psi A_{\mu}+\frac{g}{\cos\theta_{w}}\bar{\psi}\gamma^{\mu}\psi Z_{\mu} + (-\frac{\kappa}{\sqrt{2}}\bar{\ell }_{L}\psi_{R}h-\kappa'\bar{\ell }_{R}S^{\dagger }\psi_{L}+g_{S}\bar{\ell }_{R}S^{\dagger }\ell_{L}\\
    &+g_{Z}\bar{\ell }_{L}\gamma^{\mu}\psi_{L}Z_{\mu}+g_{W}\bar{\nu}\gamma^{\mu}\psi_{L}W_{\mu}^{+}+h.c.),
\end{split}
\end{eqnarray}
where
\begin{eqnarray}
\begin{split}
&g_{S}=\frac{\kappa'\kappa}{\sqrt{2}}\frac{\upsilon_{h}}{M_{F}},\\
&g_{Z}=-\frac{\kappa g}{2\sqrt{2}\cos\theta_{w}}\frac{\upsilon_{h}}{M_{F}},\\
&g_{W}=\frac{\kappa g}{2}\frac{\upsilon_{h}}{M_{F}}.
\end{split}
\end{eqnarray}
The values of the electromagnetic coupling $e$, the weak mixing angle $\theta_{w}$ and the $SU(2)_L$ coupling $g$ are taken from Particle Data Group (PDG)~\cite{ParticleDataGroup:2022pth}. We denote the VLL mass as $M_{F}$, which assumed to be equal for three generations of VLLs.

It is well known that measurements of the muon anomalous magnetic moment (AMM), $a_{\mu}=(g-2)_{\mu}/2$~\cite{Jegerlehner:2009ry,Lindner:2016bgg,Athron:2021iuf}, indicate the discrepancy from the SM prediction. The E989 experiment at Fermilab recently released an update regarding the measurement of $a_{\mu}$ from Run-2 and Run-3~\cite{Muong-2:2015xgu}. The analysis of the Muon  g-2 collaboration using the new combined value has led to $\Delta{a_{\mu}} \equiv a_{\mu}^{\text{exp}}$  $-$  $a_{\mu}^{\text{SM}} = (116592059\pm22)\times10^{-11} - (116591810\pm43)\times10^{-11} = (249\pm48)\times10^{-11}$, a discrepancy of 5.1$\sigma$ CL~\cite{Muong-2:2023cdq,Aoyama:2020ynm}. The contributions of the VLS models to the muon AMM have been carefully studied in Refs.\cite{Hiller:2020fbu,Hiller:2019tvg}, which only appear at one loop level,  are dependent on the free parameter $\kappa'$  and the masses of new particles including $M_{F}$ and $M_{S}$, and  scale quadratically with the muon mass,
\begin{eqnarray}
\begin{split}
&\Delta{a_{\mu} = \frac{\kappa'}{32\pi^{2}} \frac{m_{\mu}^{2}}{M_{F}^{2}} f(\frac{M_{S}^{2}}{M_{F}^{2}} )},
\end{split}
\end{eqnarray}
with $f(t) = (2t^{3} + 3t^{2} - 6t^{2}\ln{t} - 6t + 1)/ (t - 1)^{4}$ positive for any $t$, and $f(0)= 1$. If we demand the VLS models to solve the muon AMM anomaly,  then the coupling $\kappa'$ can be expressed in terms of the mass parameters $M_{F}$ and $M_{S}$. The contributions of the VLS models to $(g-2)_{e}$~\cite{Fan:2022eto} come from one-loop level as well and mainly depend on the coupling $\kappa$, the mass parameters $M_{F}$ and $M_{S}$, see \cite{Hiller:2019tvg,Hiller:2019mou} for details. Furthermore, the $Z\rightarrow\ell\ell$ data~\cite{ParticleDataGroup:2022pth}  constrains the mixing angles $\theta$ as $\theta \simeq\kappa v_{h}/\sqrt{2}M_{F} < \mathcal{O}(10^{-2})$. According to both AMMs and $Z\rightarrow\ell\ell$ data, we simply fix $\kappa/\kappa' = 10^{-2}$ with $\kappa' $ computed according to Eq.(5) for $M_{S}=$ 500 GeV in our numerical calculation, as done in Ref.~\cite{Bissmann:2020lge}.

\begin{figure}[H]
\begin{center}
\centering\includegraphics [scale=0.42] {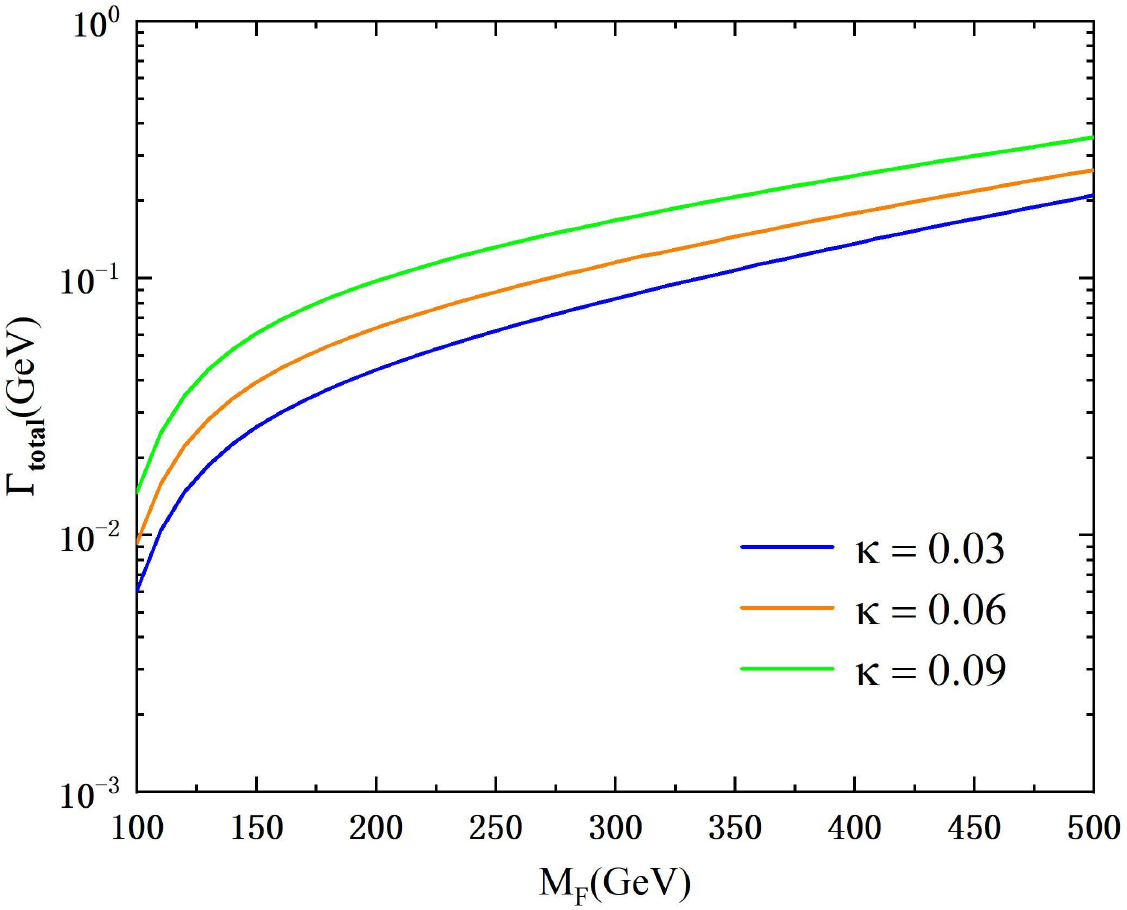}
\caption{The total decay width as function of the VLL mass $M_{F}$ for different  values of $\kappa$.}
\label{total width}
\end{center}
\end{figure}

In the singlet VLS model, the charged VLLs can decay into the final states $W\nu$, $Z\ell$, $S\ell$, $h\ell$ and the partial decay widths are
\begin{eqnarray}
\begin{split}
&\Gamma(\Psi \rightarrow W\nu)=g_{W}^{2}\frac{M_{F}}{32\pi}(1-r_{W}^{2})^{2}(2+1/r_{W}^{2}) , \\
&\Gamma(\Psi \rightarrow Z\ell)=g_{Z}^{2}\frac{M_{F}}{32\pi}(1-r_{Z}^{2})^{2}(2+1/r_{Z}^{2}),\\
&\Gamma(\Psi \rightarrow S^{\ast}\ell)=\kappa'^{2}\frac{M_{F}}{32\pi}(1-r_{S}^{2})^{2},\\
&\Gamma(\Psi \rightarrow h\ell)=\kappa^{2}\frac{M_{F}}{64\pi}(1-r_{h}^{2})^{2},\\
\end{split}
\end{eqnarray}
where $r_{X}=M_{X}/M_{F}$ for $X= W, Z, S, h$. The branching ratios as function of $M_{F}$ are given in Ref.~\cite{Bissmann:2020lge}, which show that the branching ratio $Br(\psi\rightarrow W\nu)$ is the highest one for $M_{F} \leq$ 500 GeV. So in this work, we focus on the channel of VLLs decaying into $W\nu$. To facilitate following calculations, we plot the VLL total decay width in FIG.~\ref{total width} as function of the VLL mass $M_{F}$ for  $M_{S}=$ 500GeV and different values of the coupling $\kappa$.

The VLS models satisfying asymptotic safety, possible up to the Planck scale, allow to explain the discrepancies between the SM prediction and current data of AMMs and other observables. The new particles $\psi$ and $S$ predicted by VLS models can be as light as a few hundred GeV and thus can be probed at high energy colliders. In this work, we mainly consider the possibility of detecting the singlet VLS model through pair production of  VLLs at the ILC with $\sqrt{s}=$ 1 TeV and $\mathcal{L}=$ 1 ab$^{-1}$ and only discuss the first generation of VLLs as an example.

\section{EVENT GENERATION AND NUMERICAL RESULTS}

\begin{figure}[H]
\begin{center}
\centering\includegraphics [scale=0.72] {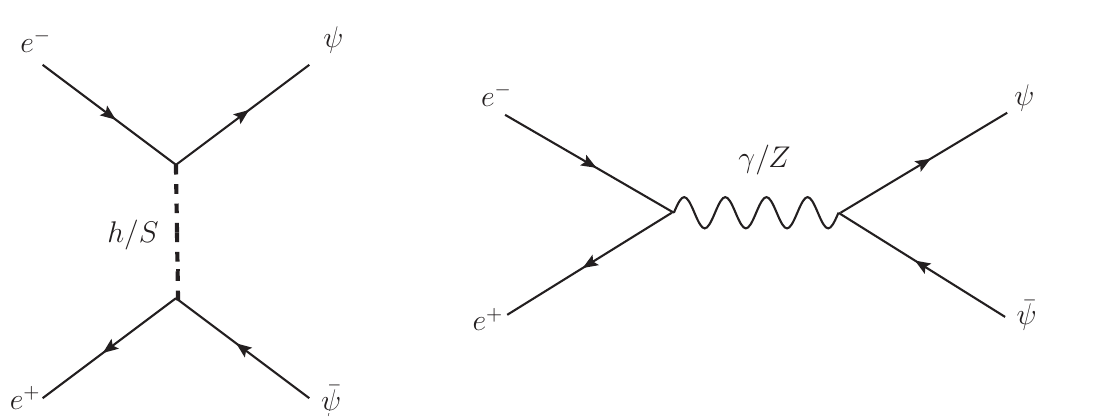}
\caption{The Feynman diagrams for the process $e^{+}e^{-}\rightarrow\psi\overline{\psi}$.}
\label{feynman diagram}
\end{center}
\end{figure}

\begin{figure}[H]
\begin{center}
\centering\includegraphics [scale=0.42] {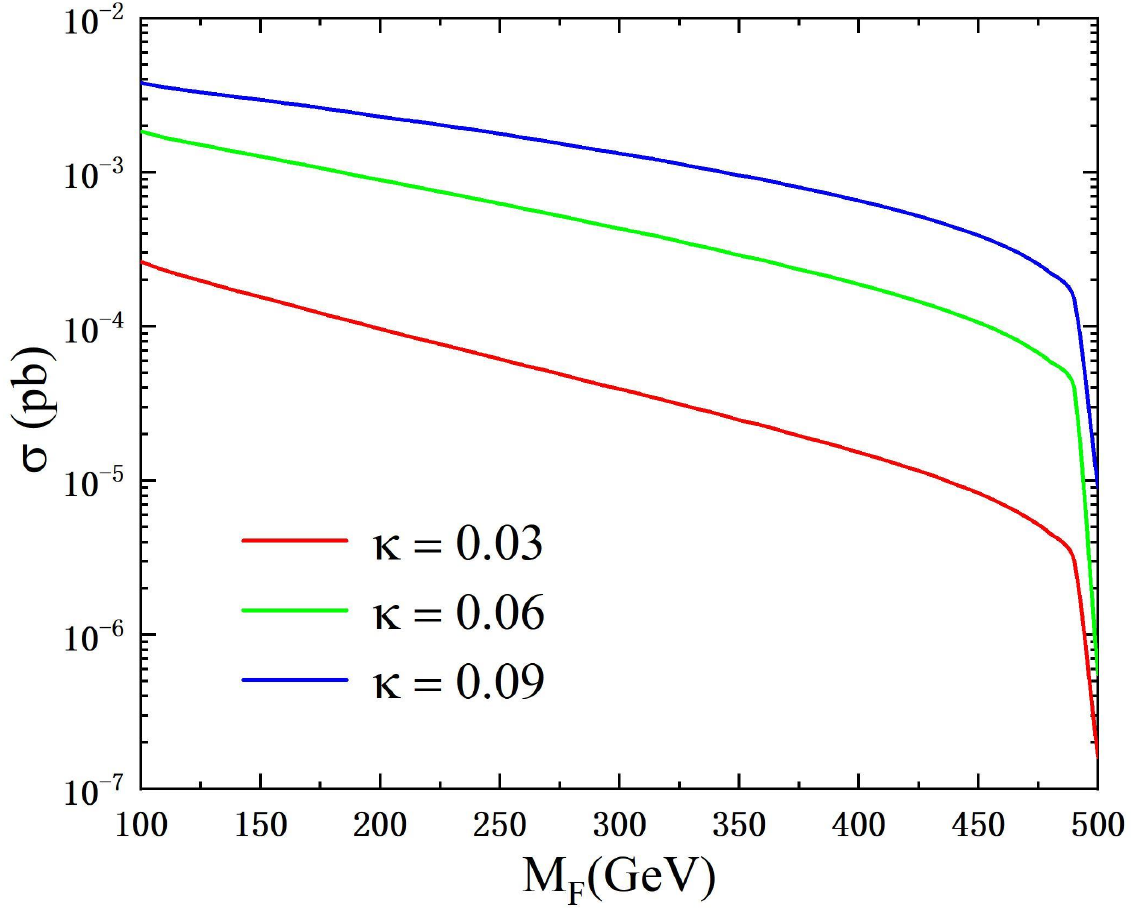}
\caption{The cross sections of the process $e^{+}e^{-}\rightarrow\psi\overline{\psi}\rightarrow W^{+}\nu_{e} W^{-}\nu_{e}\rightarrow\ell^{+}\ell^{-}\nu_{e}\bar{\nu_{e}}\nu_{\ell}\bar{\nu_{\ell}}$ as function of the VLL mass $M_{F}$ for different values of the coupling $\kappa$.}
\label{cross section}
\end{center}
\end{figure}

The Feynman diagrams for the pair production of VLLs via process $e^{+}e^{-}\rightarrow\psi\overline{\psi}$ at the ILC are shown in FIG.~\ref{feynman diagram}. We implement the singlet VLS model into \verb"FeynRules" package~\cite{Alloul:2013bka} and export the model files to the \verb"UFO" format~\cite{Degrande:2011ua}. The numerical results are given by using Monte-Carlo (MC) simulation with the \verb"MadGraph5_aMC@NLO" toolkit~\cite{Alwall:2014hca,Frederix:2018nkq}. Since the polarized $e^{+} $ beams and $e^{-} $ beams can enhance the cross section effectively, we set the polarization options as  $P_{e^{+}}=-$ 0.3, $P_{e^{-}}=$ 0.8 and assume the VLL mass in the range of 100 - 500 GeV within the detection capability of the 1 TeV ILC~\cite{Behnke:2013lya,Chakrabarty:2022dai}, the Yukawa coupling $\kappa$ in the range of 0.01 - 0.1 which is consistent with the relevant experimental data~\cite{Bissmann:2020lge}.  Considering the most large branching ratio $Br(\psi\rightarrow W\nu)$ and assuming the electroweak boson W decays into $\ell\nu$, the signal of pair production of VLLs at the ILC is a pair of charged leptons plus missing energy. The production cross sections of the signal process $e^{+}e^{-}\rightarrow\psi\overline{\psi}\rightarrow W^{+}\nu_{e} W^{-}\nu_{e}\rightarrow\ell^{+}\ell^{-}\nu_{e}\bar{\nu_{e}}\nu_{\ell}\bar{\nu_{\ell}}$ at the 1 TeV ILC are shown in FIG.~\ref{cross section} as function of the VLL mass $M_{F}$ for different values of $\kappa$. The main SM background we concerned comes from the processes $e^{+}e^{-}\rightarrow\ell^{+}\ell^{-}\nu_{\ell}\bar{\nu_{\ell}}$ and $e^{+}e^{-}\rightarrow\ell^{+}\ell^{-}\nu_{\ell}\nu_{\ell}\bar{\nu_{\ell}}\bar{\nu_{\ell}}$ with $\ell=e , \mu$. It is obvious that the signal cross section is smaller than that of the corresponding SM background (0.08931pb). The parton level events of the signal and background are required to pass through the basic cuts, so our numerical results have employed the basic cuts as follows
$$
\begin{array}{l}
	\textit{p}_{T}^{\ell}>10 \mathrm{GeV}, ~~ \quad  \Delta R _{\ell\ell}>0.4, ~~ \quad  \lvert \eta_{\ell}\rvert<2.5,\\
\end{array}
$$
where $\textit{p}_{T}$ denotes the transverse momentum, $\Delta R(x,y)=\sqrt{(\Delta\phi)^{2}+(\Delta\eta)^{2}}$ is the separation in the rapidity-azimuth plane. We transmit the parton-level events to \verb"PYTHIA8"~\cite{Sjostrand:2014zea} for showering and hadronization. Then we make a fast detector simulations by \verb"DELPHES"~\cite{deFavereau:2013fsa} using the ILD detector card. Finally, we perform the event analysis by \verb"MADANALYSIS5"~\cite{Conte:2012fm,Conte:2014zja,Conte:2018vmg}.

\begin{figure}[H]
\begin{center}
\subfigure[]{\includegraphics [scale=0.4] {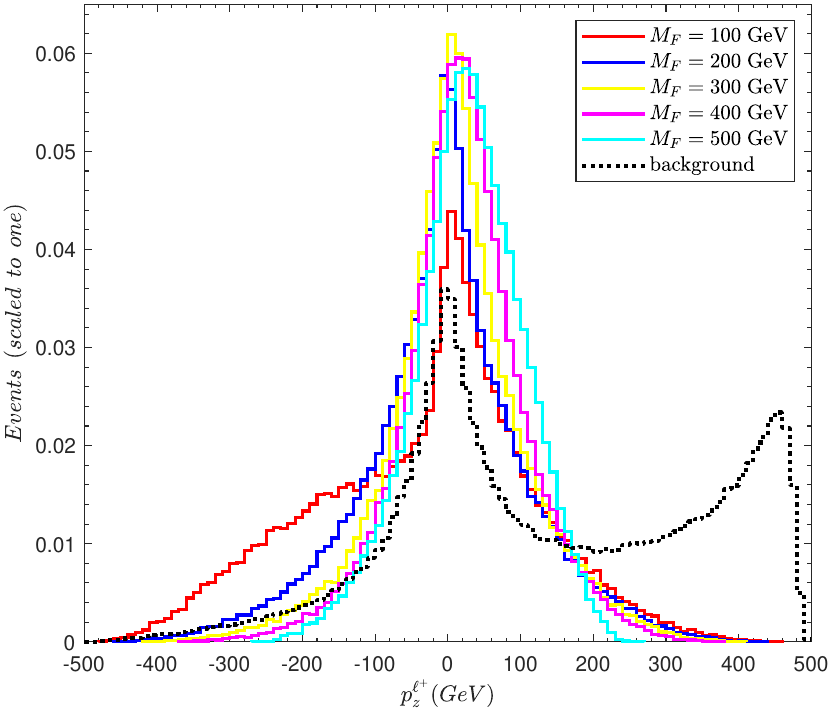}}
\hspace{0.2in}
\subfigure[]{\includegraphics [scale=0.4] {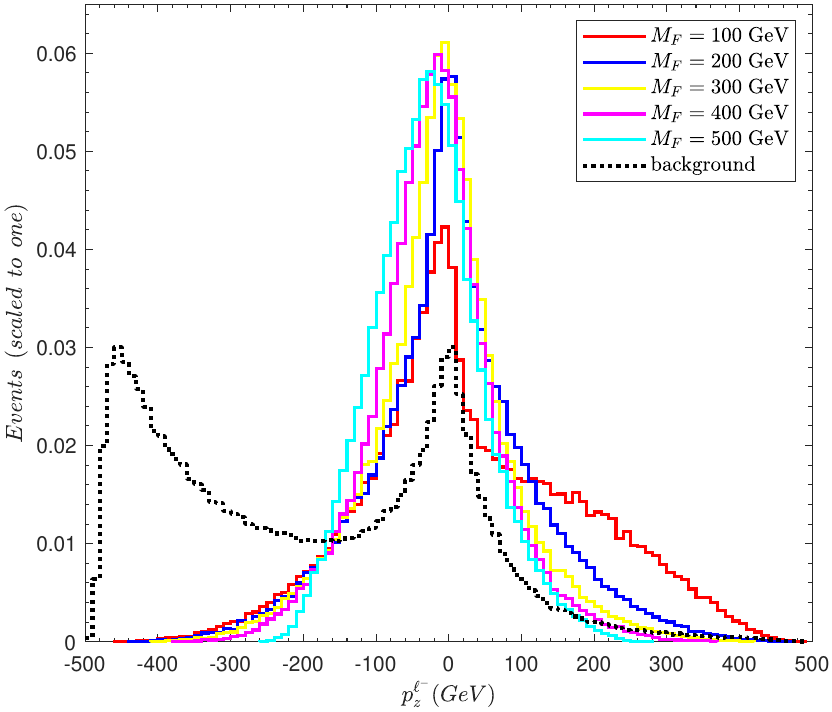}}
\hspace{0.8in}
\subfigure[]{\includegraphics [scale=0.4] {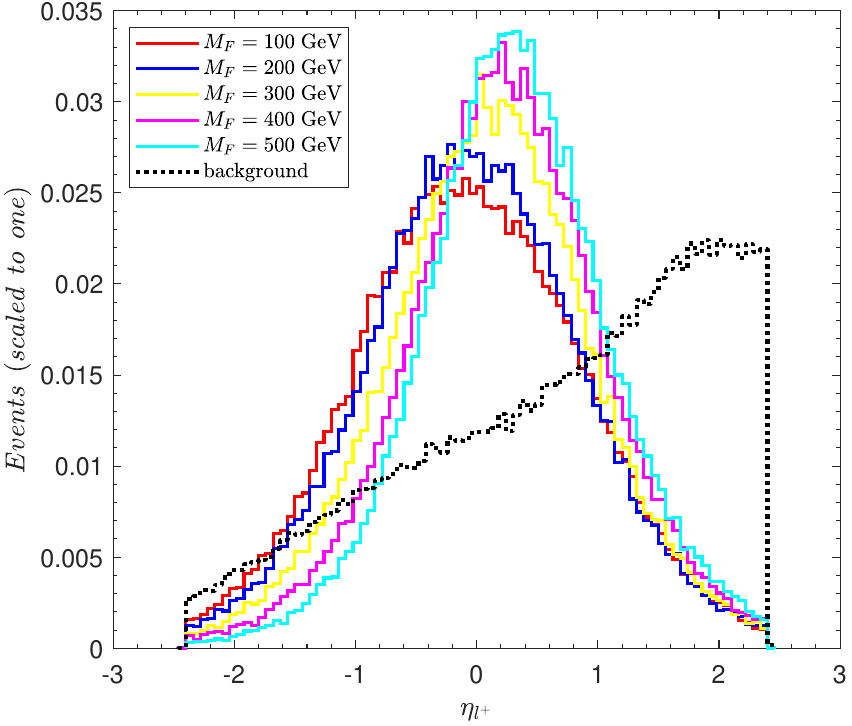}}
\hspace{0.2in}
\subfigure[]{\includegraphics [scale=0.4] {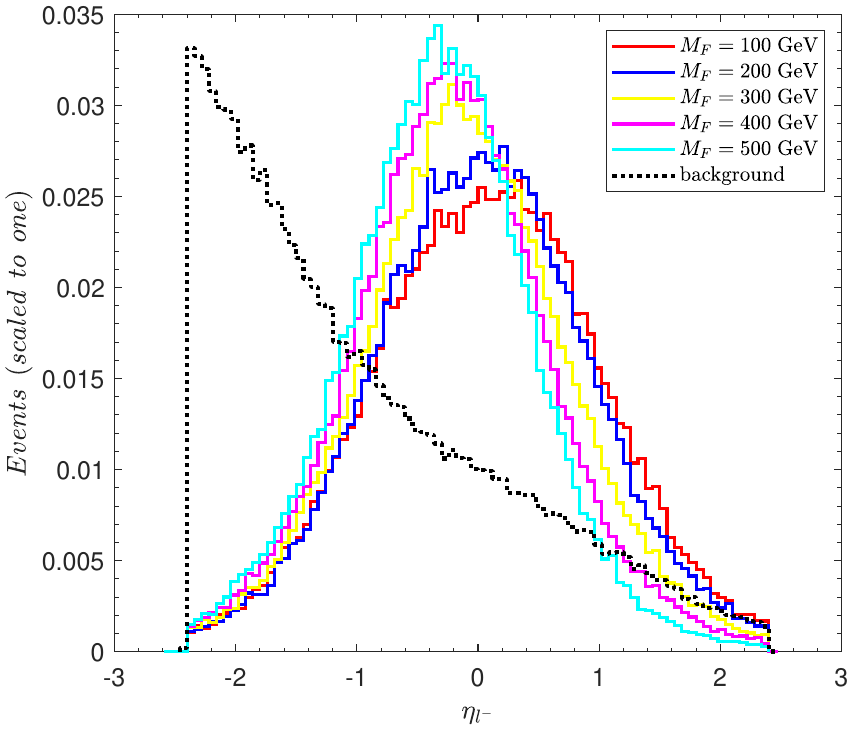}}
\caption{The normalized distributions of the observables $\textit{p}_{\textrm{z}}^{\ell^{+}}$ (a), $\textit{p}_{\textrm{z}}^{\ell^{-}}$ (b), $\eta_{\ell^{+}}$ (c) and $\eta_{\ell^{-}}$ (d) in signal and background events for various VLL masses at the 1 TeV ILC with $\mathcal{L}=$ 1 ab$^{-1}$.}
\label{distribution}
\end{center}
\end{figure}

To improve the signal and suppress the background, we serve the $\textrm{z}$-component of the momentum and the pseudorapidity of $\ell^{+}, \ell^{-}$ in the final states as effective cuts, severally. The normalized distributions of these observables are depicted in FIG.~\ref{distribution} with the solid lines for signal events and dashed lines for background events. The benchmark points are $M_{F} = $ 100 GeV, 200 GeV, 300 GeV, 400 GeV and 500 GeV selected with the assumption of $\kappa =$ 0.06, $\mathcal{L} =$ 1 ab$^{-1}$.

From FIG.~\ref{distribution}$(a)$, we can see that more signal events are distributed in the region with smaller $\textit{p}_{\textrm{z}}^{\ell^{+}}$ compared to the background. The reason is that the process $e^{+}e^{-}\rightarrow\ell^{+}\ell^{-}\nu_{\ell}\bar{\nu_{\ell}}$ dominates the background. The missing particles of background carried less energy lead to the $\textrm{z}$-component of the momentum of visible particles $\ell^{+}$ to be distributed at large momentum, whereas the momentum of the signal is mainly distributed at small masses. The $\textrm{z}$-component of the momentum of $\ell^{-}$ is similar to $\textit{p}_{\textrm{z}}^{\ell^{+}}$ as shown in  FIG.~\ref{distribution}(b). Therefore, we take the cut $\textit{p}_{\textrm{z}}^{\ell^{+}}<$ 180 GeV and $\textit{p}_{\textrm{z}}^{\ell^{-}}> -$  180 GeV, respectively. The $\eta_{\ell^{+}}$ distribution given by FIG.~\ref{distribution}(c) shows that a large proportion of the beam axis, which is significantly distinct from the background. The same goes for $\eta_{\ell^{-}}$ distribution, so we implemented the cut of $\eta_{\ell^{+}}<$ 0.8  and $\eta_{\ell^{-}}> -$ 0.8 to filter the background. Based on the characteristics of the kinematics distributions, the selected cuts are listed in table~\ref{table1} to further reduce the background.

\begin{table}[H]
\begin{center}
\caption{All cuts on the signal and background.}
\label{table1}
\begin{tabular}
[c]{c| c c c c c}
    \hline\hline
    ~~~~Basic cuts~~~~      &  $~~~~~~~\textit{p}_{T}^{\ell}>10 \mathrm{GeV}, ~~ \quad  \Delta R _{\ell\ell}>0.4, ~~ \quad  \lvert \eta_{\ell}\rvert<2.5~~~~~~~$ \\

	~~~~Cut 1~~~~      &  $~~~~~~~~~~~~~~~~~\textit{p}_{\textrm{z}}^{\ell^{+}}<180~GeV~~~~~~~~~~~~~~$ \\

	~~~~Cut 2~~~~      &  $~~~~~~~~~~~~~~~~~\textit{p}_{\textrm{z}}^{\ell^{-}}>-180~GeV~~~~~~~~~~~~~~$  \\

	~~~~Cut 3~~~~      &  $~~~~~~~~~~~~~~~~~\eta_{\ell^{+}}< 0.8~~~~~~~~~~~~~~$    \\

    ~~~~Cut 4~~~~      &  $~~~~~~~~~~~~~~~~~\eta_{\ell^{-}}> -0.8~~~~~~~~~~~~~$    \\
   \hline \hline

\end{tabular}
\end{center}
\end{table}

\begin{table}[H]\tiny
	\centering{
\caption{The cross sections of the signal and the background processes after the improved cuts applied for $\kappa=0.06$ at the 1TeV ILC with benchmark points.$~$\label{tab2}}
		\newcolumntype{C}[1]{>{\centering\let\newline\\\arraybackslash\hspace{50pt}}m{#1}}
		\begin{tabular}{m{1.5cm}<{\centering}|m{2cm}<{\centering} m{2cm}<{\centering} m{2cm}<{\centering}  m{2cm}<{\centering} m{2cm}<{\centering} m{2cm}<{\centering}}
			\hline \hline
      \multirow{2}{*}{Cuts} & \multicolumn{5}{c}{cross sections for signal (background) [pb]}\\
     \cline{2-6}
     & $M_F=100$ GeV  & $M_F=200$ GeV  & $M_F=300$ GeV & $M_F=400$ GeV &  $M_F=500$ GeV  \\ \hline
     Basic Cuts  & \makecell{$1.8351\times10^{-3}$\\$(0.08931)$} & \makecell{$8.9060\times10^{-4}$\\$(0.08931)$} &\makecell{$4.3086\times10^{-4}$\\$(0.08931)$} &\makecell{$1.8733\times10^{-4}$\\$(0.08931)$} & \makecell{$2.2202\times10^{-6}$\\$(0.08931)$}
        \\
     Cut 1  & \makecell{$1.4419\times10^{-3}$\\$(0.04564)$} & \makecell{$7.2116\times10^{-4}$\\$(0.04564)$} &\makecell{$3.5429\times10^{-4}$\\$(0.04564)$} &\makecell{$1.5573\times10^{-4}$\\$(0.04564)$} & \makecell{$1.8926\times10^{-6}$\\$(0.04564)$}
       \\
     Cut 2  &\makecell{$1.1614\times10^{-3}$\\$(0.01443)$}  & \makecell{$5.9084\times10^{-4}$\\$(0.01443)$} &\makecell{$2.9374\times10^{-4}$\\$(0.01443)$} &\makecell{$1.3048\times10^{-4}$\\$(0.01443)$} & \makecell{$1.6176\times10^{-6}$\\$(0.01443)$}
       \\
     Cut 3  & \makecell{$1.0406\times10^{-3}$\\$(0.01135)$} & \makecell{$5.2073\times10^{-4}$\\$(0.01135)$} &\makecell{$2.5001\times10^{-4}$\\$(0.01135)$} &\makecell{$1.0487\times10^{-4}$\\$(0.01135)$} & \makecell{$1.2301\times10^{-6}$\\$(0.01135)$}
     \\
     Cut 4  & \makecell{$9.9005\times10^{-4}$\\$(7.0357\times10^{-3})$} & \makecell{$4.7868\times10^{-4}$\\$(7.0357\times10^{-3})$} &\makecell{$2.1930\times10^{-4}$\\$(7.0357\times10^{-3})$} &\makecell{$8.6010\times10^{-5}$\\$(7.0357\times10^{-3})$} & \makecell{$9.3007\times10^{-5}$\\$(7.0357\times10^{-3})$}
     \\ \hline
     $SS$  & $11.04$ & $5.52$ & $2.57$ & $1.019$ & $0.011$  \\ \hline \hline
	\end{tabular}}	
\end{table}

After these improved cuts are applied, the SM background is substantially suppressed. The cross sections of the signal and background after applying the above selection cuts for the five benchmark points picked for 100 GeV $\leq{M_{F}}\leq$ 500 GeV are displayed in table~\ref{tab2} at 1 TeV ILC. Then we further calculate the statistical significance $SS=S/\sqrt{S+B}$, where $S$ and $B$ are the number of events for the signal and background, respectively. Considering the integrated luminosity of $\mathcal{L}=$ 1 ab$^{-1}$, a statistical significance of 5.521$\sigma$ (2.574$\sigma$) can be reached when $M_F$ is taken as 200 GeV (300 GeV). The significance can be greater than 5$\sigma$ for $M_F$ less than 200 GeV and will be small as $M_F$ approaches 500 GeV.

\begin{figure}[H]
\begin{center}
\centering\includegraphics [scale=0.42] {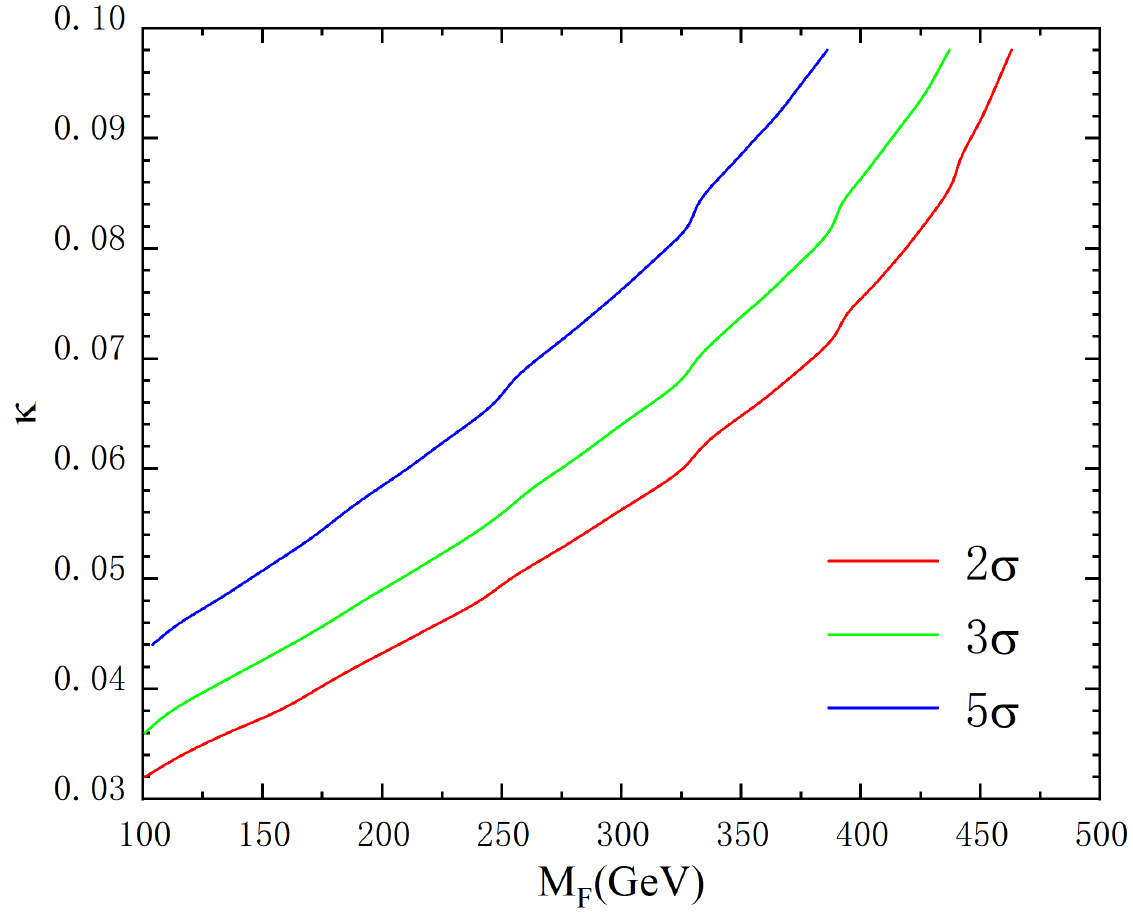}
\caption{The 2$\sigma$, 3$\sigma$ and 5$\sigma$ curves for the process $e^{+}e^{-}\rightarrow\psi\overline{\psi}$ at the ILC with $\sqrt{s}=$ 1 TeV and $\mathcal{L}=$ 1 ab$^{-1}$ in the $M _F - \kappa$ plane.}
\label{235sigma}
\end{center}
\end{figure}

In FIG.~\ref{235sigma}, the 2$\sigma$, 3$\sigma$ and 5$\sigma$ curves for the process $e^{+}e^{-}\rightarrow\psi\overline{\psi}\rightarrow W^{+}\nu_{e} W^{-}\nu_{e}\rightarrow\ell^{+}\ell^{-}\nu_{e}\bar{\nu_{e}}\nu_{\ell}\bar{\nu_{\ell}}$  at the ILC with $\sqrt{s}=$ 1 TeV and $\mathcal{L}=$ 1 ab$^{-1}$ are plotted in the plane of $M _F$-$\kappa$. As shown in FIG.~\ref{235sigma}, the Yukawa coupling $\kappa$ increases with $M_F$ increasing and breaks when $M_F$ approaches to 463 (437, 386) GeV. That is mainly because the cross section of the signal sharply decreases with increasing $M_F$ as shown in FIG.~\ref{cross section}. For the mass $M_F$ in the range of 100 - 500 GeV, the values of the statistical significance $SS$ can reach 2$\sigma$, 3$\sigma$ and 5$\sigma$ as long as the coupling coefficient $\kappa$  greater than 0.032, 0.036 and 0.044, respectively. Thus, the ILC has the potential to discover the VLLs in the considered mass range.

\section{conclusions}

Asymptotically safe extensions of the SM are one kind of interesting new physics scenarios, which predict the existence of new singlet scalars and vector-like fermions producing significant contributions to
some low-energy observables and rich collider phenomenology. The singlet and doublet VLS models are two specific models featuring three generations of either $SU(2)_L$
singlet or doublet VLLs. In this paper, we investigate the probability of probing the VLLs at the ILC with $P(e^{+}, e^{-})=(- 0.3,  0.8)$ beam polarization via the signal process $e^{+}e^{-}\rightarrow\psi\overline{\psi}\rightarrow W^{+}\nu_{e} W^{-}\nu_{e}\rightarrow\ell^{+}\ell^{-}\nu_{e}\bar{\nu_{e}}\nu_{\ell}\bar{\nu_{\ell}}$  in the framework of the singlet VLS model.  The expected sensitivities of the ILC with $\sqrt{s} =$ 1 TeV and  $\mathcal{L}$ = 1 ab$^{-1}$ to the parameter space of this kind of VLL models are derived. The results show that the parameter space $M_{F}\in$ [101, 463] GeV and $\kappa \in$ [0.032, 0.098]  can be covered by the proposed ILC.  Our conclusions are obtained for the first generation of VLLs, which are also  apply to the second generation of VLLs. Certainly, due to the rapid decay of $\tau$, making the identification very challenging, searching for the third generation VLLs at ILC needs to be further investigated, which is one of our future tasks.
\section*{ACKNOWLEDGMENT}

This work was partially supported by the National Natural Science Foundation of China under Grant No. 11875157, No. 12147214 and No. 11905093. We would like to thank Ji-Chong Yang for useful discussions.


\bibliography{VLLref}

\end{document}